# Communicating uncertainty in Indigenous sea Country monitoring with Bayesian statistics: towards more informed decision-making

Katherine Cure[§,1], Diego R Barneche[§,1,2], Martial Depczynski[1,2], Rebecca Fisher[1,2], David J Warne[3,4], James McGree[3,4], Jim Underwood[5], Frank Weisenberger[6], Elizabeth Evans-Illidge[1], Daniel Oades[8], Azton Howard[7], Phillip McCarthy[7], Damon Pyke[7], Zac Edgar[7], Rodney Maher[7], Trevor Sampi[7], and Bardi Jawi Traditional Owners[§§]

[§]Authors contributed equally to the paper

[§§]We acknowledge the contributions of the Bardi Jawi people of the Kimberley in this work. We thank them for sharing their Traditional Knowledge, gained by thousands of years of ongoing observation and connection to their sea Country

[1]Australian Institute of Marine Science, Indian Ocean Marine Research Centre, Crawley, Western Australia, Australia

[2]UWA Oceans Institute and School of Biological Sciences, University of Western Australia, Crawley, Western Australia, Australia

[3]School of Mathematical Sciences, Queensland University of Technology, Brisbane, Queensland, Australia

[4]Centre for Data Science, Queensland University of Technology, Brisbane, Queensland, Australia

[5]Gondwana Link Ltd, Albany, Western Australia, Australia

[6]Frank Weisenberger Consulting, Victoria, Australia

[7]Bardi Jawi Rangers, Kimberley Land Council, Ardyaloon, Western Australia, Australia

[8]Kimberley Land Council, Broome, Western Australia, Australia

Correspoding Author: Katherine Cure; k.cure@aims.gov.au; m: +61497849278

## Abstract

First Nations Australians have a cultural obligation to look after land and sea Country, and Indigenous-partnered science is beginning to drive socially inclusive initiatives in conservation. The Australian Institute of Marine Science has partnered with Indigenous communities in systematically collecting monitoring data to understand the natural variability of ecological communities and better inform sea Country management. Monitoring partnerships are centred around the 2-way sharing of Traditional Ecological Knowledge, training in science and technology, and developing communication products that can be accessed across the broader community. We present a case study with the Bardi Jawi Rangers in northwest Australia focusing on a 3-year co-developed and co-delivered monitoring dataset for culturally important fish in coral reef ecosystems. We show how uncertainty estimated by Bayesian statistics can be incorporated into monitoring indicators and facilitate fuller communication between scientists and First Nations partners about the limitations of monitoring to identify change.

**Introduction**

Recognition of the profound cultural and spiritual connections of First Nations peoples with their sea Country and the benefits of their inclusion in adaptive ocean management is building momentum worldwide (Lauer and Aswani 2010; Artelle et al. 2019; Houde et al. 2022). Despite a millennial history of marine stewardship using traditional ways, Indigenous peoples inclusion in government-led ocean management has been mostly limited, partly because the conservation and management landscape is biased to western science and governance structures that are foreign to them (Ross et al. 2009; Peer et al. 2022). Practices are often limited to tick-box-type exercises of stakeholder engagement (Strand et al. 2022), with limited opportunity to genuinely influence or co-design the decision-making process (Smit et al. 2022). The United Nations Declaration on the Rights of Indigenous Peoples (UNDRIP) established a universal framework to uphold the rights and interests of First

Peoples including their role in management and governance of land and sea Country. Governments and researchers are increasingly creating spaces for Indigenous roles and perspectives to improve our collective understanding and management of natural resources (Souther et al 2023; Nakashima et al 2018). For example, the Australian Institute of Marine Science (AIMS) has established an Indigenous Partnership Plan, Policy and Project Team to facilitate collaboration between marine scientists, Indigenous people and Traditional Owners. Key to the success of these efforts are genuine two-way partnerships between Indigenous peoples and science organisations that are inclusive of Traditional Ecological Knowledge (TEK), can provide training, and enable data collection and reporting appropriate for linking science to policy making (Depczynski et al. 2019; Souther et al. 2023).

Australia's first peoples, the Aboriginal and Torres Strait Islander Peoples, represent the oldest continuous culture on earth (Malaspinas et al 2016). For over 50,000 years they have established deep spiritual and cultural connections to Country of the Australian continent and adjacent seas. As Traditional Owners (TO) and custodians of Australia's land and sea Country, their rights and interests include the cultural responsibility to look after Country and safeguard it for future generations (Rist et al. 2019). This has resulted in vast holdings of TEK based on detailed observation and experimentation, transmitted between generations through cultural expressions and traditions, encompassing climate shifts and major sea level changes to coastline and island systems (Nunn and Reid 2016; Horstman and Wightman 2001). With this immense body of knowledge, Indigenous Australians have managed Australia's marine ecosystems for tens of thousands of years (Allen and O'Connell 2003; Nunn and Reid 2016), effectively protecting biodiversity and culturally significant sites, and providing a secure food source for their people. Currently however, even the most remote marine ecosystems are under increasing pressure from climate change, habitat loss, fisheries, and tourism (Wilson et al. 2006; Graham et al. 2008; Wilson et al. 2012; Jones et al. 2018). These new challenges require cooperation between Indigenous peoples, science, and government agencies, to improve understanding of marine and coastal ecosystems and improve management and social outcomes (Ross et al. 2009; Dobbs et al. 2016; Rist et al. 2019).

The role of Indigenous peoples in managing land and sea Country in Australia and actively participating in its governance, has been re-asserted in the past three decades by a combination of joint management arrangements in Marine Protected Areas (MPAs), designation of Indigenous Protected Areas (IPAs), and funding of over 170 community based Indigenous Ranger groups employing over 1900 rangers to support conservation (Ross et al. 2009; Rist et al. 2019). IPAs have been established in Indigenous-owned land or sea and currently make up about 50% of Australia's National Reserve System. They empower TOs with official recognition and resources for governance, ranger employment, and operational funds.  IPAs are established through a formal agreement with government to promote conservation of biodiversity and cultural resources, and development of a management plan, which sets out how TOs propose to look after land and sea Country. Since the first Healthy Country Plan (HCP) was developed by Wunambal Gaambera TOs in the Northern Kimberley, many other TO groups in Australia chose the Healthy Country Planning methodology, an adaptation of the widely used Open Standards for the Practice of Conservation, to develop IPA management plans (Conservation Measures Partnership 2020). HCPs outline key targets for conservation and set out strategies to abate threats, restore targets, and evaluate their ongoing health and impacts through performance indicators. Science partnerships are essential elements of IPA management plans for the collection of systematic monitoring data to inform, implement and evaluate management actions (Rist et al. 2019).

The Australian Institute of Marine Science (AIMS), a nationally sponsored scientific research agency tasked with conducting marine monitoring as a core priority. Since 2018, AIMS has been working in partnership with Indigenous communities across northern Australia to co-design monitoring programs and inform sea Country management. Within this program,

AIMS has partnered with the Bardi Jawi and Oorany Rangers (BJR) in the remote Kimberley region of northwest Australia to monitor coral reefs and fish populations in the Bardi Jawi IPA. Using modern science and technologies underpinned by TEK, the partnership has co-designed a cross-cultural monitoring program that targets key indicators in their Management Plan (Bardi Jawi Niimidiman Aborginal Corporation 2012) to inform marine management strategies.

Monitoring is the repeated sampling over space and time to collect baseline data and evaluate spatial and temporal variation so that reliable estimates of change can be obtained to inform reactive management decisions (Magurran et al. 2010; Nash and Graham 2016). Monitoring provides information on the impact of pressures to marine environments and can identify trends and vulnerabilities to improve decision making before cultural, ecological, or social values of an area become significantly degraded (Emslie et al. 2008; Babcock et al. 2010). To be impactful, particularly in Indigenous-partnered science and marine management contexts, monitoring trends need to be effectively communicated between scientists, TOs, community and decision makers, an area that proves particularly challenging because of the different streams of knowledge (Strand et al. 2022). Moreover, statistical trends carry uncertainty in their estimates depending on the monitoring spatiotemporal design and analysis models, and such uncertainty needs to be carefully reported on and considered to provide transparent pathways to decision making and demonstrate a realistic evaluation of western science to sensitively detect changes for TOs.

Bayesian statistics offer a powerful way to model spatiotemporal trends and their uncertainty. Given some prior information combined with information from the data, posterior distributions depict probabilistic estimates of model parameters (Kruschke and Liddell 2018). These distributions offer a highly intuitive and visual approach for communicating statistical outcomes to broad audiences. For fish populations, which exhibit high levels of abundance variation in space and time (Holbrook et al. 1994; Anderson and Millar 2004; Cure et al. 2018), providing probabilistic information on population trends is essential to inform more rapid decision making in an adaptive management setting. Consider, for example, a hypothetical scenario where changes in fish abundance over time define indicators which are then used to categorise the health status of a population. In Bayesian statistics, the posterior distribution of fish abundance yields not only the central estimate (e.g., changes in mean or median) but also the probability (a.k.a. credibility) of that abundance estimate encompassing a target health status category. Therefore, health indicators calculated from Bayesian models provide a direct and more interpretable measurement of the uncertainty and credibility of detected changes between sampling events.

Here, we present a case study using Bayesian multilevel models to assess changes in fish populations of traditional and recreational importance in coral reef habitats within the Bardi Jawi IPA. We showcase how this framework has improved the reporting of results and increased the impact of monitoring data through a more direct alignment with the reporting format of IPA management and monitoring plans (Bardi Jawi Niimidiman Aboriginal Corporation 2012; Bardi Jawi Rangers 2020). We conclude our work by discussing how we envision this approach evolving further and in a manner that can be adopted and adapted by coastal Indigenous communities and their partners across Australia and the globe.

**Materials and Methods**

*Monitoring Co-design*
The Bardi Jawi Native Title Determination includes over 204,000 hectares of sea Country and 200 km of coastline along the Dampier Peninsula in northwest Australia (Fig. 1a). An IPA was established in 2013, with the vison of maintaining healthy land and sea country, as well as traditional cultural knowledge and practice for future generations (Bardi Jawi Niimidiman Aborginal Corporation 2012). Under the direction of their elders, the BJR, who are also Traditional Owners of Bardi Jawi Country, are responsible for implementing the IPA Management Plan—hereafter referred to as the HCP—as well as monitoring the health of its key conservation targets and reporting outcomes back to community. Overlapping the Bardi Jawi IPA is the Bardi Jawi Gaarra Marine Park, a newly appointed joint management marine park that has been co-designed between government and TOs, including the BJR.

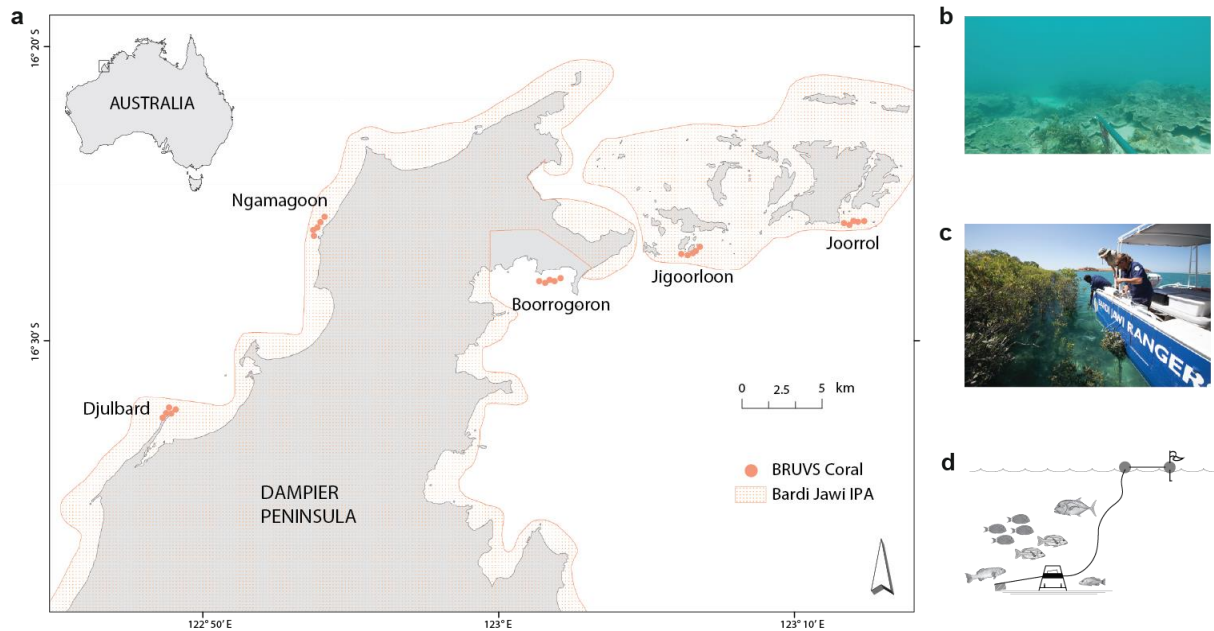

**Fig. 1.** Map of the Bardi Jawi IPA monitoring sampling locations for *aarli* (fish) in coral reef habitat (a). Side panel shows an example of coral reef habitat (b), as well as the Bardi Jawi Rangers and AIMS staff deploying BRUVS (c), and a schematic diagram of BRUVS in the seabed (d).

Bardi Jawi sea Country is rich in biodiversity, with high levels of endemism and a mosaic of habitat types including inter-tidal pools, mangroves, seagrass, algal beds, and well-developed coral reef systems (Fox and Beckley 2005; Thorburn et al. 2007). Tidal fluctuations in this region are one of the largest in the world, reaching up to 12 m and creating tidal currents of up to 10 knots (Purcell 2002; Lowe et al. 2015). Bardi and Jawi people are saltwater people, custodians of their sea Country and have historically deep and strong biophysical knowledge and connections to the sea, on which they depend for their livelihood and food security. Fish (*aarli* in Bardi Jawi language) are a particularly abundant and readily accessible resource for Bardi Jawi people.

In 2018, AIMS and the BJR ran a workshop to co-design a marine monitoring program for the Bardi Jawi IPA, merging expertise of scientists at AIMS in coral reef and fish monitoring, the TEK held by rangers and elders, and key conservation targets in the management plan (*marnany*—coral reef, and *aarli*—fish). We used a participatory mapping approach (Davies et al. 2020) to collect TEK on local habitats and species and establish habitat maps for the area. We then used this knowledge and maps to select monitoring sites and design a monitoring program that considered ranger working capacity, included sites within different clan areas, accounted for different levels of human access, and targeted areas of known habitat of culturally important species and resources (Depczynski et al 2019). During this workshop, we also provided training to rangers about quantitative scientific monitoring, sampling techniques and technologies, with an emphasis on the use of long-term monitoring data as a tool for detecting change, reporting to government and management agencies, and participating in policy change.

*Sampling Fish Populations*

To sample fish populations (*aarli*) in Bardi Jawi sea Country, we used Baited Remote Underwater Video Stations (BRUVS), a non-destructive diver-less method (Cappo et al. 2003; Whitmarsh et al. 2017) which alleviates the need for expert fish identification in the field and ensures a long-term record of the fish community at each location. This method is also the safest option to sample fish populations in the region because of its high energy tidal currents and the abundance of sharks and crocodiles.

BRUVS were deployed annually since 2018, during August-September targeting neap tides and a window of most favourable wind and swell conditions. Because environmental conditions vary drastically along the coastal Kimberley region, selection of consistent deployment times is crucial for minimising variation in currents and visibility, which could influence fish counts. All samples were taken in full daylight to avoid the effects of crepuscular times in fish behaviour (Helfman 1986), despite some variation in time of day being inevitable given tide selection is prioritised and tide times vary considerably from year to year.

Samples were collected at five sites spanning the western and eastern sides of the Dampier Peninsula (Fig. 1a). One site, Boorrogoron, was only sampled on a single year (2020); this site was an addition to the original sampling design, included to foster partnerships with the adjacent Kimberley Marine Research Station. Five 30-minute deployments, separated by a minimum distance of 250 m, were undertaken at each site in coral reef habitat for a total of 20 BRUVS samples each year (25 samples in 2020, when Boorrogoron was surveyed). Annual sampling targeted the same GPS coordinates for BRUVS to minimise spatial variation over time. Monitoring started using single-camera systems, and in 2020 moved to a twin-camera stereo system to enable size measurements. BRUVS cameras (GoPro Hero5 Black, 30 frames per second, 1920 x 1080-pixel resolution) were placed on a lightweight stainless-steel frame separated by 380 mm and faced a bait bag filled with 1 kg of crushed pilchards (*Sardinops sagax*).

Imagery from BRUVS were analysed using EventMeasure software (www.seagis.com.au) to determine fish species diversity and abundance as *MaxN*, a relative measure of abundance (maximum number of individuals from each species viewed at a single still frame during each video sample; Ellis and DeMartini 1995; Willis and Babcock 2000). All fish in BRUVS videos were identified to the lowest taxonomic level possible and treated as species complexes in cases where identification based on video imagery was not possible (i.e., *Plectropomus* spp., coral trout complex). *MaxN* estimates were then extracted for each of the ten species or groups of species selected as important indicators by the BJR for their importance in traditional and recreational fisheries (Table 1). Data presented on this case study focuses on this abundant subset of the fish community.

**Table 1.** Ten *aarli* (fish) species or group as species identified as important indicators of the health of populations important for food security in Bardi Jawi sea Country. This list includes species important to both Indigenous and recreational fisheries.

| Bardi Jawi name | Common name | Species or Group |
| --- | --- | --- |
| Barrambarr | Bluebone | *Choerodon schoenleinii* |
| Barrbal | Rabbitfish | *Siganus lineatus* |
| Biidib | Rock cods | *Epinephelus* spp. |
| Biindarral | Coral trout | *Plectropomus* spp. |
| Gambarl | Surgeonfish | *Acanthurus grammoptilus* |
| Goolan | Bluespot tuskfish | *Choerodon cyanodus* |
| Irrariny | Grass Emperor | *Lethrinus laticaudis* |
| Jirral | Trevallies | *Carangoides* spp., *Caranx* spp., *Gnathanodon* spp. |
| Jooloo | Stripey Snapper | *Lutjanus carponotatus* |
| Maarrarn | Mangrove Jack | *Lutjanus argentimaculatus* |

*Healthy Country Plan*

Fish (*aarli*) are one of seven culturally important targets within the Bardi Jawi IPA HCP with a goal of restoring health to these targets identified by TOs as the most important to be looked after (Bardi Jawi Niimidiman Aboriginal Corporation 2012). Bardi Jawi people have concerns that fish are threatened by increased recreational fishing and want to make sure that this food source continues to be available now and for future generations. Their aspiration is to improve the health of this resource through time.

Through their healthy country planning, Bardi Jawi people have also developed a plan to monitor the management outcomes in their HCP (Bardi Jawi Rangers 2020). This plan includes a set of indicators used to monitor the health of management targets and document their status based on a traffic light system: red—*poor*—restoration is very difficult and may result in extinction, yellow—*fair*—outside acceptable range of variation and requires human intervention, light green—*good*—within acceptable range of variation and some intervention required for maintenance, and dark green—*very good*—most desirable status and requires little intervention for maintenance. Adjustments to management are then made based on these assessments to restore indicator health where needed (Fig. 2).

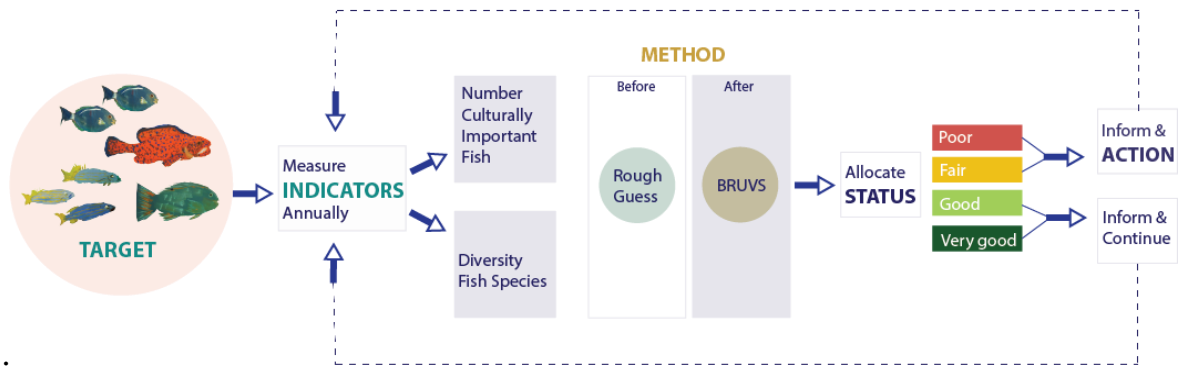

.

**Fig. 2.** Diagram detailing the process for evaluation of management targets as part of the Bardi Jawi Healthy Country Plan. Here we show the process for evaluating the health of *aarli* (fish) based on two indicators, which are currently evaluated via BRUVS, following partnership between rangers and scientists. Fish in this image were painted by Bardi Jawi children during an on-Country workshop to share monitoring results with community.

For fish, monitoring indicators are the number of culturally important fish, and the diversity of fish species; this case study focuses on the former. Before working in partnership with AIMS, these indicators were evaluated based on a qualitative estimate. Currently, indicators are evaluated using BRUVS as a method for quantifying fish abundance and diversity, and results compared annually to a 2018 baseline corresponding to the first sampling year.

As part of designing a monitoring plan, the BJR and AIMS have made every effort to include sites along the extent of Bardi Jawi sea Country, so that the various clan groups can be informed as to what is occurring in their respective areas. However, the HCP is an overall plan for all sites collectively.

*Statistical Analyses*
<u>*Main model.*</u> The purpose of this analysis is to illustrate a pipeline from modelling to communication rather than to model a particular species population trend. For that reason, we sum the *MaxN* across the ten fish species in Table 1—hereafter simply *A*. We model *A* following key assumptions that were primarily guided by the Bardi Jawi HCP, which envisions monitoring key indicators of fish populations for the entire Bardi Jawi sea Country, rather than specific sites. In that realisation, we assumed that: 1) sites are a random representation of an overall Bardi Jawi mean fish abundance; 2) the first year of monitoring (2018) is considered the baseline year from which the status of fish abundance health is calculated for subsequent years (2019–2020); 3) the period between 2018–2020 represents background natural variation in the absence of any known disturbance, and therefore years are a random representation which should depict natural background variation in fish abundance; 4) BRUVS are considered to be spatially fixed among years.

Fish abundance (integer counts) data, *A*, often exhibit high spatiotemporal variation (Sale 1978; Sale and Douglas 1984; Holbrook et al. 1994; Anderson and Millar 2004; Cure et al. 2018) which can lead to over-dispersion in the data (Fig. S1). Therefore, we assume that the data are generated by a Negative Binomial process, $NB$:

$$A \sim NB(\mu, \varphi)$$

$$\ln(\mu) = \beta_0 + \Delta_B + \Delta_S + \Delta_Y + \Delta_{S:Y}$$

$$\beta_0 \sim \mathcal{N}(0,1);$$

$$\Delta_B = \zeta_B \sigma_{\Delta_B}; \; \zeta_B \sim \mathcal{N}(0,1); \; \sigma_{\Delta_B} \sim \Gamma(2,2);$$

$$\Delta_S = \zeta_S \sigma_{\Delta_S}; \; \zeta_S \sim \mathcal{N}(0,1); \; \sigma_{\Delta_S} \sim \Gamma(2,2);$$

$$\Delta_Y = \zeta_Y \sigma_{\Delta_Y}; \; \zeta_Y \sim \mathcal{N}(0,1); \; \sigma_{\Delta_Y} \sim \Gamma(2,2);$$

(1)

$$\Delta_{S:Y} = \zeta_{S:Y}\sigma_{\Delta_{S:Y}}; \zeta_{S:Y} \sim \mathcal{N}(0,1); \sigma_{\Delta_{S:Y}} \sim \Gamma(2,2);$$

$$\varphi \sim \Gamma(2,1),$$

where $\beta_0$ is the "global" Bardi Jawi among-sites and among-years mean fish abundance on the natural log scale, $\Delta_{[B,S,Y,S:Y]}$ are respectively BRUVS-, sites-, year- and site-year-specific deviations from $\beta_0$, and $\varphi$ is the over-dispersion parameter. All $\Delta_*$ parameters were estimated indirectly as the multiplication of the standardised effects $\zeta_*$ and their respective standard deviations $\sigma_*$. Priors were weakly informative, as calibrated by prior predictive checks, so that the conclusions are largely data driven. The prior sampling distributions are the Gaussian ($\mathcal{N}$(mean, standard deviation)), and Gamma ($\Gamma$(shape, inverse scale)). We include the site-year hierarchical interaction term, $\Delta_{S:Y}$, to account for any site-specific temporal idiosyncrasies, and because this allows us to calculate site- and year-specific fish abundance means in addition to the overall mean $\beta_0$ originally included as an indicator in the HCP. Importantly, these hierarchical effects also allow us to directly derive indicators from year-specific posterior distributions. For example, $e^{\beta_0+\Delta_{Y=2019}} / e^{\beta_0+\Delta_{Y=2018}}$ provide a full posterior distribution of 2019-to-baseline fish abundance ratios that can be directly mapped to the health status categories of the HCP (Fig. 3). Moreover, different areas of the posterior distribution, each corresponding to a different category, can be integrated to yield a status credibility (Fig. 3). For example, in Fig. 3 the status of the fish abundance is most likely *good*, although there is some possibility it is either *fair* or *very good*. Status of *poor* has a low credibility. The major advantage of this approach is communication: probabilities are intuitive and easier to convey to a general audience.

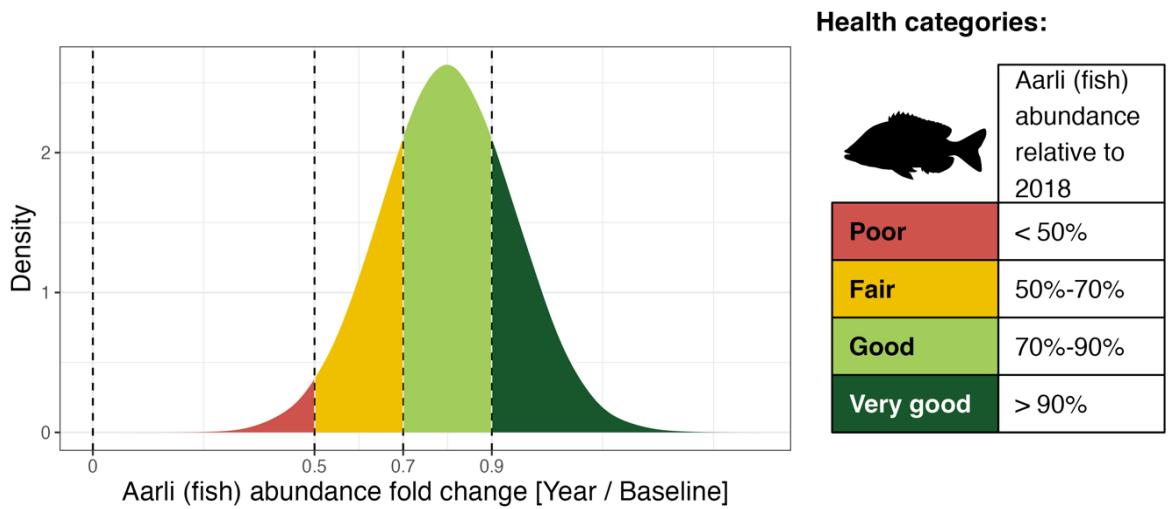

**Fig. 3.** Hypothetical posterior distribution depicting the probability of fish abundance being of a particular HCP health status category (different colours). Fold changes are calculated from the posterior distribution of year-specific estimates of fish abundance, as explained in the main text. The panel on the right shows the health categories assigned by the Traditional Owners (TOs) in the Healthy Country Plan (HCP).

The dataset contained 64 observations of fish abundance on coral reef habitat, $A$ (imagery for one BRUVS was not suitable for analyses). The model was fitted under a Bayesian framework using the package brms version 2.18.0 (Bürkner 2017) in R version 4.1.2 (R Core Team 2021) to determine posterior distributions and associated 95% highest posterior density intervals (HDI) for the fitted parameters. The posterior distributions of model parameters (Table 2) were estimated using No-U-Turn Sampler (NUTS) Hamiltonian Monte Carlo (HMC) by constructing four chains of 5,000 steps each. Half of these iterations (2,500) were used as a warm-up, so a total of 10,000 steps were retained to estimate posterior distributions (i.e., 4 × (5,000 - 2,500) = 10,000). All four independent chains reached convergence, i.e., the Gelman-Rubin statistic (Gelman and Rubin 1992), $\hat{R}$, was approximately 1 for all parameters. We adopted a target average proposal acceptance probability of 0.99, and a maximum tree depth of 20, i.e., the maximum number of steps in each iteration was $2^{20}$. No divergent transitions were observed. We also calculated a Bayesian $R^2$ (Gelman et al. 2019). Posterior predictive checks to assess goodness-of-fit are provided in the online Supplementary Information (Fig. S2).

**Table 2.** Model estimates using Bayesian methods. Parameter names correspond to those in equation 1. Lower and upper 95% highest density intervals as well as standard deviation were calculated from posterior distributions. The Gelman-Rubin statistic (Gelman and Rubin 1992), $\hat{R}$, shows that all four chains have converged.

| Parameter | Mean estimate | S.D. | L-95% HDI | U-95% HDI | $\hat{R}$ |
|---|---|---|---|---|---|
| $e^{\beta_0}$ | 10.60 | 5.74 | 0.83 | 20.90 | 1 |
| $\sigma_{\Delta_S}$ | 0.22 | 0.12 | 0.02 | 0.45 | 1 |
| $\sigma_{\Delta_Y}$ | 0.47 | 0.35 | 0.01 | 1.13 | 1 |
| $\sigma_{\Delta_{S:Y}}$ | 0.59 | 0.21 | 0.21 | 1.02 | 1 |
| $\sigma_{\Delta_B}$ | 0.71 | 0.57 | 0.01 | 1.87 | 1 |
| $\varphi$ | 1.96 | 0.43 | 1.18 | 2.79 | 1 |

*Evaluation of monitoring design.* One of the goals of the monitoring partnership was to evaluate the effectiveness of the current design in terms of detecting changes in fish abundance over time. We did this by running a power simulation considering increasing sampling effort ($\alpha$ = {5, 10, 20} BRUVS, but keeping the number of sites fixed) and different levels of multiplicative declines relative to baseline ($\rho$ = {0.05, 0.25, 0.5, 0.7, 0.9, 1}) in a new year. We employed 500 draws from the existing posterior distributions of model parameters in equation (1) to simulate datasets with 3 years of non-disturbance in fish abundance (similar to original data), and a new year where a decline effect is applied to the mean baseline Bardi Jawi fish abundance, e.g., $e^{\beta_0 + \Delta_{Y=2018}} \times \rho$. The approach can be formalised in four steps:

Step 1: Simulate non-disturbed data between 2018–2020

$$A'_1, A'_2, \ldots, A'_{64} \sim NB(\mu, \varphi).$$

Step 2: Simulate disturbed data for a new year with a $j^{th}$ decline relative to baseline, $\rho$, and a $k^{th}$ sampling effort, $\alpha$.

$$\Delta^*_Y \sim \mathcal{N}(0, \sigma_{\Delta_Y})$$

$$\Delta^*_{S:Y} \sim \mathcal{N}(0, \sigma_{\Delta_{S:Y}})$$

$$\Delta^*_{B,\alpha_k} \sim \mathcal{N}(0, \sigma_{\Delta_B})$$

$$\ln(\mu^*_{j,k}) = \ln(e^{\beta_0 + \Delta_{Y=2018}} \rho_j) + \Delta_S + \Delta^*_Y + \Delta^*_{S:Y} + \Delta^*_{B,\alpha_k}$$

$$A^* \sim NB(\mu^*_{j,k}, \varphi).$$

It is important to note that BRUVS-attributable deviations, $\Delta^*_{B,\alpha_k}$, were only simulated for the additional BRUVS when $\alpha = \{10, 20\}$, i.e., we only simulated 5 and 15 values of $\Delta^*_{B,\alpha_k}$, respectively. The original 5 $\Delta_B$ estimated in equation (1) were used in all scenarios assuming the location of the original 5 BRUVS remained constant.

Step 3: Concatenate simulated responses, $A'' = \{A', A^*\}$.

Step 4: Evaluate a modified version of model in equation 1 which has an added dummy vector, $X$, to the linear predictor representing "before" (0, 2018–2020) and "after" (1, new year) states,

$$A'' \sim NB(\nu, \varphi)$$

$$\ln(\nu) = \ln(\mu) + \beta_1 X.$$

This four-step approach allows to test for the probability of $\beta_1$ being negative, i.e., $E[I((\beta_1|D) < 0)]$ (i.e., the hypothesis, where $D$ is the data). Steps 1–4 were repeated 500 times for each combination of BRUVS sampling effort (3 levels) and declines (6 levels), each for a different posterior draw from parameters in equation (1), totalling 500 × 18 = 9,000 simulated datasets and model runs. For each scenario, we then computed the average probability of $\beta_1$ being negative across 500 simulations as a measure of statistical power. Moreover, at each iteration we evaluated the expectation $\rho = e^{\beta_0 + \Delta_{Y=new\ year}} / e^{\beta_0 + \Delta_{Y=2018}}$, i.e., the expectation of getting the simulated HCP category from the estimated decline, by integrating the different regions of its posterior distribution in relation to the HCP. Model fitting specifications follow the same as those described above for equation (1).

*Sharing Results on-Country.* As part of the annual monitoring work with the BJR, a time was set aside on-Country to share and discuss results—*yarning*. This time is crucial for allowing cross-cultural ecological understanding, developing relationships, and testing different ways to communicate science results (Davies et al. 2020). Discussions were centred around informative presentation formats, simplified explanations of underlying data analyses, and what we could understand from current results regarding changes in fish community metrics. Complementary to this, communications were also extended to the broader Bardi Jawi community, including Elders and decision makers, children via a one-week workshop with the local school, presentations to the Bardi Jawi Steering Committee, production of short films, and co-presentations between rangers and scientists at national science conferences. During these activities spanning the three years of the program, scientists kept detailed notes on the level of engagement and feedback received, to qualitatively evaluate how the results from this monitoring program have been understood, interpreted, and accepted by the BJR and other members of community. Feedback was used to develop the format and informational content of science communication products for the monitoring program (see Discussion).

**Results**

*Main model*

Our statistical model explained 38% of the variation observed in the data (Bayesian R² 95% highest posterior density interval (HDI): 13%–64%). The mean combined *MaxN* of the 10 important fish species/groups—$e^{\beta_0}$—in Bardi Jawi country was 10.6 fish across all sites and years, although this estimate was quite uncertain (HDI: 0.83–20.9; Fig. 4; Table 2). This uncertainty was also reflected in how much *MaxN* varied across space and time: 1.6-fold (i.e., $e^{2\times\sigma_{\Delta_B}}$; HDI: 1.0–2.4) across BRUVS, 4-fold (i.e., $e^{2\times\sigma_{\Delta_S}}$; HDI: 1.0–9.6) across sites, 19.9-fold (i.e., $e^{2\times\sigma_{\Delta_Y}}$; HDI: 1.0–41.7) across years, and 3.6-fold (i.e., $e^{2\times\sigma_{\Delta_{S:Y}}}$; HDI: 1.3–7.0) across sites and years. The data exhibited a moderate degree of over-dispersion (mean $\varphi$ = 1.96; HDI: 1.18–2.79).

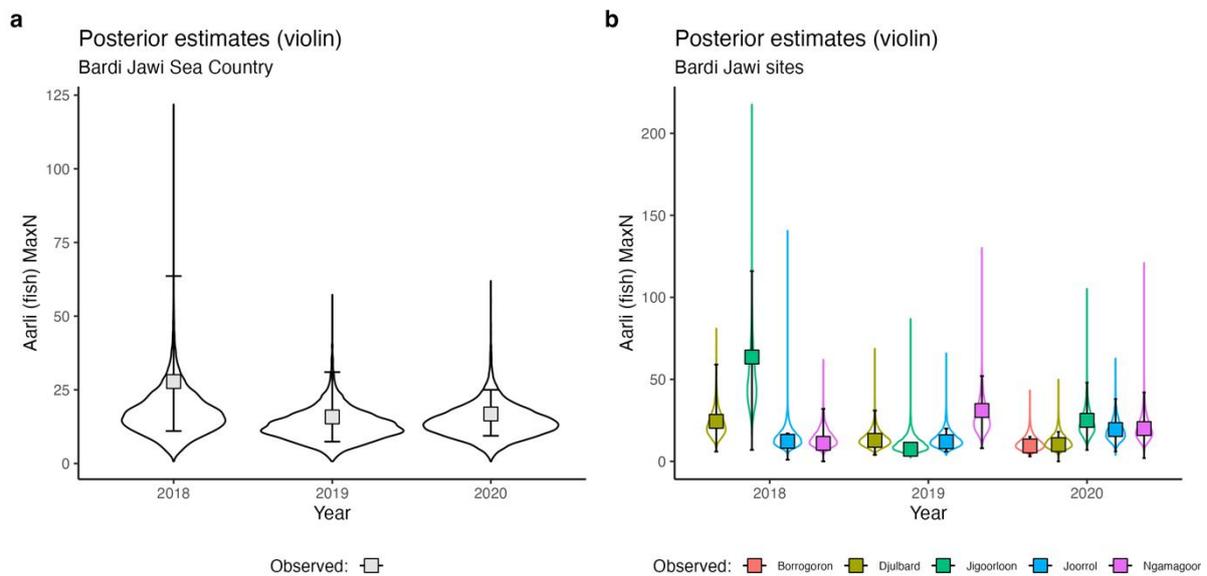

**Fig. 4.** *MaxN* combined across the 10 important fish species/groups in Bardi Jawi Country. (a) shows the mean trends per year for Bardi Jawi sea Country, whereas (b) shows trends per site. Symbols and error bars (95% HDI) were calculated from data, whereas violin plots depict posterior distributions estimated by the model described in equation (1).

Overall, the sum of *MaxN* from the 10 important fish species/groups mostly declined in 2019 and 2020 for Bardi Jawi sea Country combined (Fig. 4a). Based on the posterior distributions of fold change relative to the baseline year 2018 for Bardi Jawi sea Country, there was a 75% and 67% chance the fish abundance had declined (the proportion of the posterior below 1) in 2019 and 2020 respectively (Fig. 5a). This decline was observed for most sites, except for Ngamagoon which increased substantially in 2019, then declined back to baseline levels in 2020 (Fig. 4b). These trends were translated in varied health status across sites and years (Fig. 5b). For example, while sites Djulbard and Jigoorloon exhibited mostly a *poor* or *fair* status in both years, Ngamagoon and Joorrol exhibited high credibility for a *very good* status (Fig. 5b).

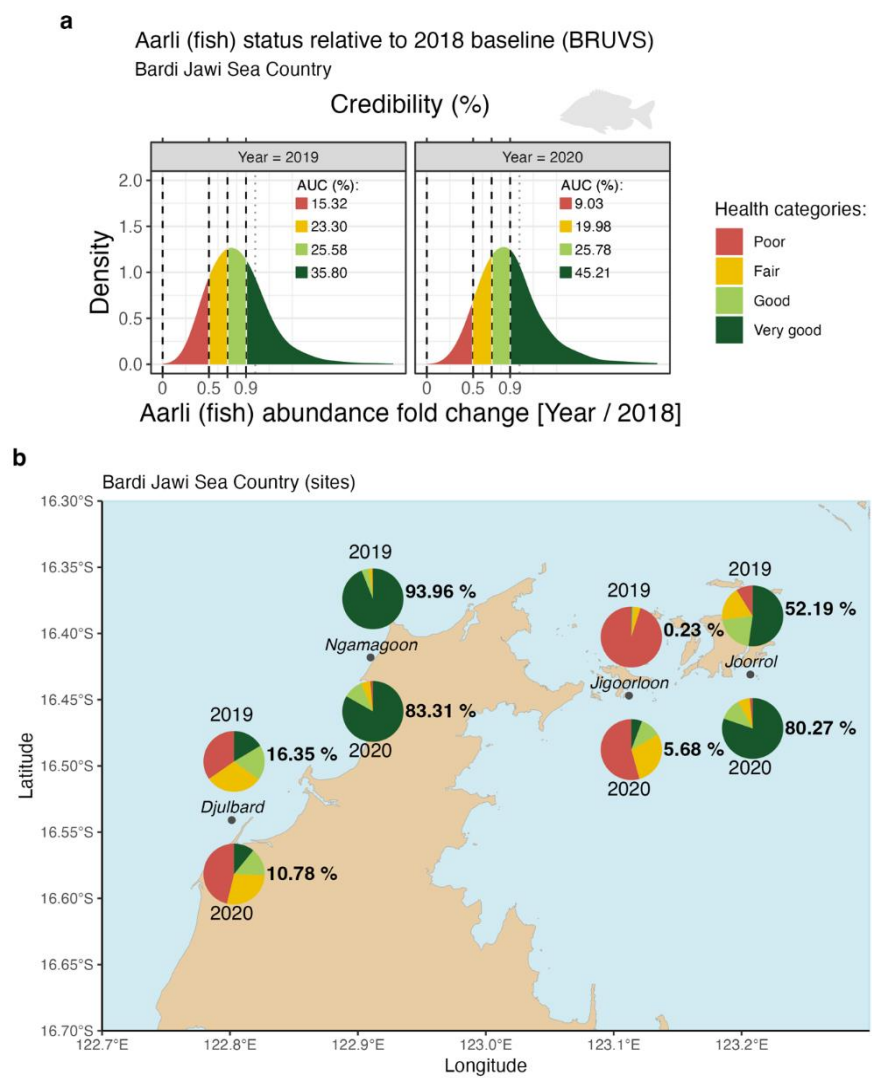

**Figure 5.** Overall (a), and site-specific (b) health status of *MaxN* combined across the 10 most important fish species/groups in Bardi Jawi sea Country. Health status classifications follow the health categories in the Bardi Jawi HCP. Density plots in (a) depict the credibility of each category

based on posterior distributions of abundance fold change ($e^{\beta_0 + \Delta_{Y=\{2019,2020\}}} / e^{\beta_0 + \Delta_{Y=2018}}$), with credibility being calculated as the relative area under the curve (AUC; in %). The vertical grey dotted line at 1 indicates no change. In (b), credibility estimates were re-expressed as pie charts for site-specific calculations on the map, i.e., incorporating $\Delta_S + \Delta_{S \cdot Y}$ into the calculations. Bold values depict credibility of *MaxN* being *very good* at each site and year. The site Boorrogoron is not displayed because it was only sampled in 2020 and therefore, we could not back-calculate changes relative to the baseline year, 2018.

*Evaluation of monitoring design*

Our simulation approach revealed limited capacity of the current monitoring design to detect immediate change over the following monitoring year, regardless of the sampling effort adopted (Fig. 6a). Specifically, an >80% credibility to detect change was only observed once 95% of the fish population were removed from 2018 baseline model estimates, and more moderate declines such as 30% yielded a credibility of 50%, thus the overall confidence was equally split between decline and increase in fish population abundance. Moreover, the simulation that imposed no change to fish abundance relative to baseline (0% decline in Fig. 6a) yielded 37.5% credibility of a negative decline across the different efforts. However, the expectation for a simulation that recovers a true (simulated) mean of 0% change, on average, should be 50%, i.e., half of the combined posterior distributions should be negative and the other half positive.

The uncertainty in the posterior distributions combined across the different simulations resulted in large uncertainty for the different HCP health categories (Fig. 6b). For example, contrary to expectation (bottom colour bar in Fig. 6b), simulated declines of 10–50% did not translate into high credibility of *fair* and *good* status, however their credibility should have been higher if the modelled simulated data (Step 4 in the Methods section for Evaluation of monitoring design) yielded parameter values close to the original simulation process (Steps 1–3 in the Methods section for Evaluation of monitoring design).

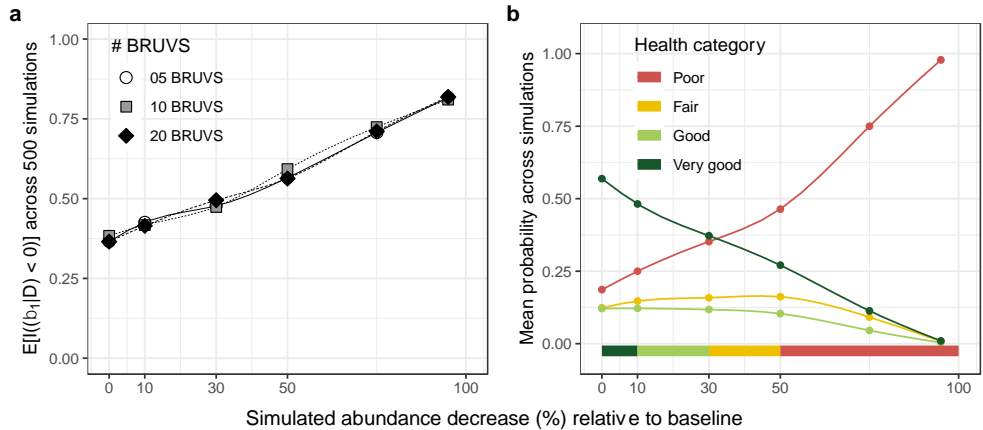

Detecting change in Aarli (fish) abundance

**Fig. 6.** Results of simulations used to evaluate monitoring design. (**a**) shows the average probability (combined across 500 Bayesian simulations) to detect a negative decline in fish abundance at a new year (see Evaluation of monitoring design in the Methods section for a full description of the approach) for three different sampling efforts (i.e., number of BRUVS). (**b**) shows the mean probabilities of each HCP indicator category (coloured points and lines) across the simulated population decline gradient. The x axis depicts the different values of $\rho$—re-expressed as a percentage decline—used for the simulations. Different sections of the x axis (horizontal colour bar at the bottom) encompass the Healthy Country Plan indicator categories. The mean probabilities of each category were calculated from posterior distributions of $e^{\beta_0+\Delta_{Y=new\ year}} / e^{\beta_0+\Delta_{Y=2018}}$. Given the lack of obvious difference among sampling effort scenarios (**a**), (**b**) was drawn from 20-BRUVS scenarios only. For visual purposes, connecting lines in (**a**) and (**b**) have been smoothed using splines in all panels.

## Discussion

Here we explored data collected by the co-designed environmental monitoring partnership between the Bardi Jawi people and AIMS. We specifically explored a Bayesian framework approach to estimate uncertainty and help break the communication barrier between western science and Indigenous people when reporting statistical monitoring trends, by: (1) providing an indication of the degree of uncertainty in the change estimates calculated in the results, and (2) facilitating a better understanding of the limitations of monitoring for identifying sudden small changes in fish community metrics. We argue this approach addresses this objective by qualifying estimated changes in monitored environmental parameters (e.g., fish population abundance) with a numeric probability statement that aligns with a simple traffic-light system to best inform decision-making processes of sea Country managers (Budescu et al. 1988). Moreover, the flexibility of multilevel models in a Bayesian framework allows for direct calculation of metrics which, in this case—changes in abundance relative to the baseline year of monitoring—correspond directly to the health of environmental monitoring indicators (HCP health categories). We first discuss our overall results and modelling limitations, and then turn to potential general implications of our work to Indigenous monitoring and better ways that western science can simultaneously learn from and contribute to it.

*Main model*

Our model captured only 38% of the variation in the data purely via hierarchical structure. This may be improved in the future by the addition of measuring environmental covariates that could help explain more of the data in a deterministic way (e.g. temperature, chlorophyl-a (proxy for productivity), habitat from the BRUVS field of view). It is also unclear whether the Negative Binomial distribution, which was empirically determined based on the nature of the data, is the true underlying distribution that governs fish abundance in space and time. Characterising and determining fish abundance dynamics has been a topic of much debate for decades (Sale 1978; Sale and Douglas 1984; Anderson and Millar 2004; Irigoyen et al. 2013; Thibaut and Connolly 2020). In future, we anticipate further improvements to covariates measured and our data modelling capacity to increase the data variability explained in the models and provide more accurate results with greater confidence and less uncertainty.

*Evaluation of monitoring design*

It was surprising that the modelled probability of the current monitoring design to detect change was only better than chance odds (50/50) only once a large proportion of the fish population was removed. Although these estimates result from a model with several assumptions and based on a limited time series, our inability to detect a useful effect size with reasonable certainty raises at least three questions for this monitoring program going forwards.

First, is power to detect change limited by the length of our time series? Three years of data may not sufficiently encompass the spatiotemporal variation of fish communities in the highly variable coastal environment of the Kimberley. Longer datasets may be required to properly account for and override natural variation in space and time, so that trends in abundance and the effects of disturbance, are more accurately identified (Magurran et al. 2010; Lindenmayer and Cunningham 2011). Identifying the minimum number of monitoring years for this to occur is essential information for the design of any monitoring program and setting expectations of what western science approaches can and cannot do in the service of marine resource management.

Second, could the power to detect change with greater certainty be improved by increasing the sampling effort? Our results indicate that increasing the number of BRUVS by up to four times at a particular site, will have no impact on improving the power to detect change. This means that a larger number of samples would not reduce variability in our estimates of fish abundance. There are two potential reasons for this. Firstly, our indicator metric (*MaxN10*) includes species with different lifestyles (site-attached benthic species, roving schools, and pelagics) and different levels of attraction to bait (herbivores and carnivores), so that an assessment may need to be made separately for each of these groups. Secondly, high environmental variation results in highly variable fish communities at very small spatial scales (250 to 500 m). Care was taken to minimize variation by being as consistent as possible in BRUVS location, time of year, time of day, tide, and visibility. However, these factors may vary drastically in the Kimberley even within the bounds of our efforts for consistency and could be preventing us from detecting smaller declines in abundance.

Third, where do we go from here? The aim of the Bardi Jawi people is to reduce threats to culturally important fish species and improve their abundance to guarantee food security for their community. Therefore, for this monitoring program to be fit-for-purpose, we need to further explore options to improve the power to detect change in fish abundance through time with confidence. This may involve modelling fish abundance with metrics other than *MaxN10* (Schobernd et al. 2014; Sherman et al. 2018), collecting more fine scale environmental data using loggers attached to BRUVS, re-thinking how change is evaluated (i.e., comparison to baseline vs. moving mean), or expanding the spatial extent of the current sampling design. The latter is currently being considered with the recent creation of the Bardi Jawi Gaara Marine Park co-managed by Bardi Jawi and the Western Australian state government. An adaptive monitoring approach informed by the current dataset to amend the current sampling design may be possible without losing the integrity of the current dataset (Lindenmayer et al. 2011). Workshopping this approach together with researchers, TOs and state government park managers will be key to its success.

*Decision making with uncertainty*

Communication of monitoring results to non-scientific audiences involves sharing a simple message on the health status of a resource. In most cases, this involves presenting temporal change as a trend—increase, decrease, no change—or, in this example, as a health category. This simple message is then used by decision makers and managers to implement strategic changes or maintain status quo. In most cases, this simplification of message requires scientists to exclude associated uncertainty. However, understanding this uncertainty is key to decision-making (Pople et al. 2007).

We have found that providing Bayesian probability estimates for all health categories in the Bardi Jawi HCP, offers an intuitive way to communicate both uncertainty and confidence in estimates, providing a stronger base for more accurate decision-making from monitoring data (see Fig. 7). Importantly, by including spatial hierarchies in the model, we can further explore these probabilities at the site level and focus action where it is most needed. In this example, the indicated decline at sites Djulbard and Jigoorloon should be closely watched.

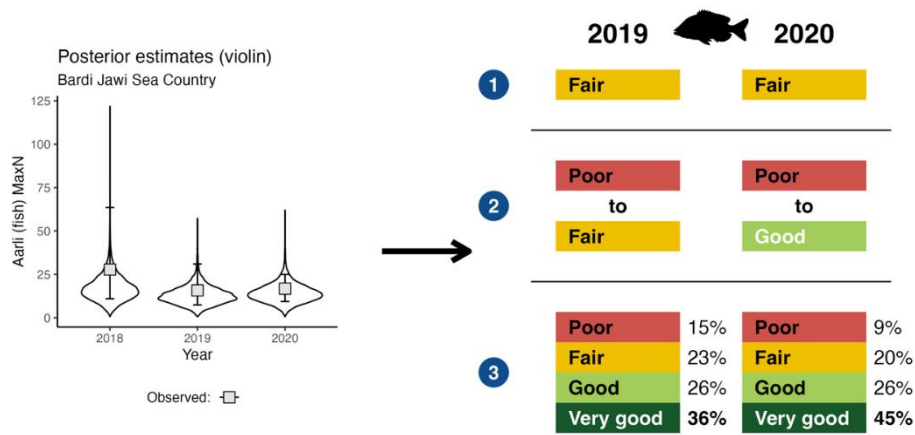

**Fig. 7.** Representation of three possible communication options from monitoring data collected in the Bardi Jawi Rangers-AIMS monitoring partnership. The plot on the left is redrawn from Fig. 4a and used as an example. On the right, 1) takes only the mean value into account, and yield 57% and 60% of the abundance relative to the 2018 baseline; therefore, *fair* would be the health category reported for both 2019 and 2020; 2) introduces uncertainty associated with the mean estimate (based on the observed HDIs), so that a series of health categories could be possible, and 3) presents Bayesian estimates of credibility (% of AUC) for each possible category in the Healthy Country Plan (see Fig. 5a). We argue that option 3 is more intuitive, provides added credibility and is better suited to inform decision-making.

Better understanding of the uncertainty associated with our long-term estimates of fish abundance, has also led to re-thinking the appropriateness of existing health categories for evaluating resource status. For example, although the aim of the Bardi Jawi people is to improve fish abundance, the best *very good* category still allows for a 10% decline in abundance. Furthermore, simulations of population decline gradients are not being reliably identified and placed into the appropriate HCP health category. It may be more appropriate to adjust the ranges used to place *aarli* (fish) in the HCP categories, to include increasing trends (>100% of baseline) and more reliably identify population decline gradients triggering a management response. These are evolving questions that will require refining over time in collaboration with rangers and TOs, with current results guiding us into the future.

*Moving forward*

Despite the uncertainties in rapidly and sensitively detecting changes in the local fish community, the Bardi Jawi community have nonetheless embraced the monitoring program and commitment to the collaboration remains strong. The results presented here are part of an important journey of understanding for both parties. For the western science practitioners of AIMS, it provides an opportunity to understand where sampling improvements can be made and how to best communicate monitoring results to a non-scientific community. For TOs, it highlights that western science has its own limitations, that science is a journey of discovery rather than a definitive destination. For both, it represents an opportunity to come together, combine traditional ecological knowledge with western science, and collaborate to improve the program. There is strong understanding from both knowledge bases that ecosystems are inherently under constant change and that the reasons for this is often complicated and not easily understood. On-going visits to Country to re-engage across all members of community is integral to this broadening of sea Country knowledge and its future management.

Explaining the complex mathematics behind Bayesian modelling is a more difficult concept. Rangers and community are still more open to seeing results presented as annual means connected via trendlines. When talking to these graphs during meetings and presentations, they focus on the mean value and the trend it shows with respect to previous years. However, presentation of probability density plots, especially when comparing between years and dividing the area under the curve into different colours using a traffic light system has been tested with the BJR and is generally seen as intuitive. Nonetheless, the focus of conversations and feedback from the rangers is the strong need for selection of a single major colour graphics (i.e., traffic light system) showing resource status that can be easily shared with community. Further on-Country *yarning* to better communicate the need for understanding uncertainty and credibility and linking this to decision-making is planned for future work.

For the time being, the monitoring program has been a sensational learning experience for all. The community seems engaged, informed, and empowered. In fact, known monitoring result outputs were used in the marine park planning of the new Bardi Jawi Gaarra Marine Park. No decision-making process regarding resource status or harvest quotas for traditional fisheries has yet occurred because of this work. The Bardi Jawi people have a long history of managing their fisheries using traditional methods which mostly involve seasonal catch restrictions, with catch being limited to when fish are fat (Rouja et al. 2003). However, with increasing pressures to their sea Country including from increased recreational fishing, monitoring data may play an increasingly important role into the future to guide new management policies for both the IPA and the marine park.

Long-term ecological monitoring is key to evidence-based environmental policy, decision making and management. Keeping track of resources over long timeframes is already understood and deeply embedded in Aboriginal and Torres Strait Islander's TEK. It has formed the basis for an intricate knowledge system to understand natural patterns based on long-term observations of when resources can be harvested. With current changes in increasing population, climate change, species extinction and restoration activities amongst others, long-term evidence-based monitoring is key to evaluating effects of change and developing ecologically sustainable resource management strategies that promote ecological and social wellbeing. We envision that the lessons learnt from this work will guide TOs, scientists and managers, into better designed and fit-for-purpose monitoring work that can help support these goals. Therefore, we will end by providing the following guidelines to assist in developing a robust joint monitoring partnership that values all ecological knowledge: 1) manage expectations – neither western science or TEK has all the knowledge and answers but together they are complementary, 2) set out clear intentions of what is being sought from a monitoring program and associated data from the beginning of the partnership, 3) evaluate the sampling plan early with available data and revise/adapt if necessary, 4) discuss uncertainty in long-term estimates regularly and from the beginning, 5) include several indicators for the health status of a resource so as not to rely on a single metric such as *MaxN10*, ideally including Traditional metrics and data on resource use, and 6) be flexible to adaption in monitoring plans, effort required and designs, to improve its utility to marine management and conservation outcomes.

**Acknowledgements.** We thank the Bardi Jawi people of the Kimberley for welcoming us on Country, and for their continued participation in this work. We remain deeply respectful of their trust in our partnership and committed to arriving at better two-ways monitoring work. Thanks also to the AIMS leadership team for believing in this work and helping establish NAMMA (North Australian Marine Monitoring Alliance), the WAMSI Kimberley Indigenous Saltwater Science Project (KISSP) for enabling the start of a relationship with the BJR, and the Kimberley Marine Research Station for supporting work at Boorrogoron. DJW and JMM acknowledge support from the Centre for Data Science at the Queensland University of Technology. DJW acknowledges support from the ARC Centre of Excellence for Mathematical and Statistical Frontiers (ACEMS; CE140100049) Industry Collaboration Support Scheme. We also thank Julie Vercelloni and Kerrie Mengersen for suggestions on earlier drafts of this manuscript. High performance computing resources for the sampling design experiments were provided by the eResearch Office at the Queensland University of Technology.

**References**

Allen, J., and J. F. O'Connell. 2003. The long and the short of it: Archaeological approaches to determining when humans first colonised Australia and New Guinea. *Australian Archaeology* 57: 5–19. doi:10.1080/03122417.2003.11681758.

Anderson, M. J., and R. B. Millar. 2004. Spatial variation and effects of habitat on temperate reef fish assemblages in northeastern New Zealand. *Journal of Experimental Marine Biology and Ecology* 305: 191–221. doi:10.1016/j.jembe.2003.12.011.

Artelle, K. A., M. Zurba, J. Bhattacharrya, D. E. Chan, K. Brown, J. Housty, and F. Moola. 2019. Supporting resurgent Indigenous-led governance: A nascent mechanism for just and effective conservation. *Biological Conservation* 240. Elsevier: 108284. doi:10.1016/j.biocon.2019.108284.

Babcock, R. C., N. T. Shears, a C. Alcala, N. S. Barrett, G. J. Edgar, K. D. Lafferty, T. R. McClanahan, and G. R. Russ. 2010. Decadal trends in marine reserves reveal differential rates of change in direct and indirect effects. *Proceedings of the National Academy of Sciences of the United States of America* 107: 18256–61. doi:10.1073/pnas.0908012107.

Bardi Jawi Niimidiman Aborginal Corporation. 2012. Bardi Jawi Indigenous Protected Area Management Plan 2013-2023. 52 pp. One Arm Point, Kimberley.

Bardi Jawi Rangers. 2020. *Bardi Jawi IPA Plan of Management Monitoring Plan*. 39 pp. One Arm Point, Kimberley.

Budescu, D. V., S. Weinberg, and T. S. Wallsten. 1988. Decisions based on numerically and verbally expressed uncertainties. *Journal of Experimental Psychology: Human Perception and Performance* 14: 281–294.

Bürkner, P.-C. 2017. Brms: An r package for bayesian multilevel models using stan. *Journal of Statistical Software* 80: 1–28.

Cappo, M., Euan. Harvey, H. Malcolm, and P. Speare. 2003. Potential of video techniques to monitor diversity, abundance and size of fish in studies of marine protected areas. In *Aquatic protected areas-What works best and how do we know? Proc World Congr on Aquat Protected Areas*, ed. J. Beumer, A. Grant, and D. Smith, 455–464. Australian Society for Fish Biology, North Beach, Western Australia.

Conservation Measures Partnership. 2020. Open Standards for the Practice of Conservation, Version 4.0|2020. 80 pp.

Cure, K., J.-P. A. Hobbs, T. J. Langlois, D. V. Fairclough, E. C. Thillainath, and E. S. Harvey. 2018. Spatiotemporal patterns of abundance and ecological requirements of a labrid's juveniles reveal conditions for establishment success and range shift capacity. *Journal of Experimental Marine Biology and Ecology* 500. doi:10.1016/j.jembe.2017.12.006.

Davies, H. N., J. Gould, R. K. Hovey, B. Radford, G. A. Kendrick, T. A. Land, S. Rangers, and C. D. Mcgonigle. 2020. Mapping the Marine Environment Through a Cross-Cultural Collaboration 7: 1–15. doi:10.3389/fmars.2020.00716.

Depczynski, M., K. Cook, K. Cure, H. Davies, L. Evans-Illidge, T. Forester, K. George, J. Gould, et al. 2019. Marine monitoring of Australia's indigenous sea country using remote technologies. *The Journal of Ocean Technology* 14: 60–75.

Dobbs, R. J., C. L. Davies, M. L. Walker, N. E. Pettit, B. J. Pusey, P. G. Close, Y. Akune, N. Walsham, et al. 2016. Collaborative research partnerships inform monitoring and management of aquatic ecosystems by Indigenous rangers. *Reviews in Fish Biology and Fisheries* 26. Springer International Publishing: 711–725. doi:10.1007/s11160-015-9401-2.

Ellis, D., and E. DeMartini. 1995. Evaluation of a video camera technique for indexing abundances of juvenile pink snapper, Pristipomoides filamentosus, and other Hawaiian insular shelf fishes. *Oceanographic Literature Review* 9: 786.

Emslie, M. J., A. J. Cheal, H. Sweatman, and S. Delean. 2008. Recovery from disturbance of coral and reef fish communities on the Great Barrier Reef, Australia. *Marine Ecology Progress Series* 371: 177–190. doi:10.3354/meps07657.

Fox, N. J., and L. E. Beckley. 2005. Priority areas for conservation of Western Australian coastal fishes: A comparison of hotspot, biogeographical and complementarity approaches. *Biological Conservation* 125: 399–410. doi:10.1016/j.biocon.2005.02.006.

Gelman, A., and D. B. Rubin. 1992. Inference from iterative simulation using multiple sequences. *Statistical Science* 7: 457–472.

Gelman, A., B. Goodrich, J. Gabry, and A. Vehtari. 2019. R-squared for bayesian regression models. *The American Statistician* 73: 307–309.

Graham, N. A. J., T. R. McClanahan, M. A. MacNeil, S. K. Wilson, N. V. C. Polunin, S. Jennings, P. Chabanet, S. Clark, et al. 2008. Climate warming, marine protected areas and the ocean-scale integrity of coral reef ecosystems. *PloS one* 3: e3039. doi:10.1371/journal.pone.0003039.

Helfman, G. 1986. Fish behaviour by day, night and twilight. In *Behavior of teleost fishes*, ed. T. Pitcher, 479–512. London: Chapman & Hall.

Holbrook, S. J., M. J. Kingsford, R. J. Schmitt, and J. S. Stephens Jr. 1994. Spatial and Temporal Variation in Assemblages of Temperate Reef Fish. *American Zoologits* 75: 26–30.

Horstman, M., and G. Wightman. 2001. Karparti ecology: Recognition of Aboriginal ecological knowledge and its application to management in north-western Australia. *Ecological Management and Restoration* 2: 99–109. doi:10.1046/j.1442-8903.2001.00073.x.

Houde, M., E. M. Krümmel, T. Mustonen, J. Brammer, T. M. Brown, J. Chételat, P. E. Dahl, R. Dietz, et al. 2022. Contributions and perspectives of Indigenous Peoples to the study of mercury in the Arctic. *Science of the Total Environment* 841: 156566. doi:10.1016/j.scitotenv.2022.156566.

Irigoyen, A. J., D. E. Galván, L. A. Venerus, and A. M. Parma. 2013. Variability in Abundance of Temperate Reef Fishes Estimated by Visual Census. *PLoS ONE* 8: e61072. doi:10.1371/journal.pone.0061072.

Jones, K. R., C. J. Klein, B. S. Halpern, O. Venter, H. Grantham, C. D. Kuempel, N. Shumway, A. M. Friedlander, et al. 2018. The Location and Protection Status of Earth's Diminishing Marine Wilderness. *Current Biology* 28. Elsevier Ltd.: 2506-2512.e3. doi:10.1016/j.cub.2018.06.010.

Kruschke, J. K., and T. M. Liddell. 2018. The Bayesian New Statistics: Hypothesis testing, estimation, meta-analysis, and power analysis from a Bayesian perspective. *Psychonomic Bulletin & Review* 25: 178–206. doi:10.3758/s13423-016-1221-4.

Lauer, M., and S. Aswani. 2010. Indigenous knowledge and long-term ecological change: Detection, interpretation, and responses to changing ecological conditions in pacific island communities. *Environmental Management* 45: 985–997. doi:10.1007/s00267-010-9471-9.

Lindenmayer, D. B., and R. B. Cunningham. 2011. Longitudinal patterns in bird reporting rates in a threatened ecosystem: Is change regionally consistent? *Biological Conservation* 144: 430–440. doi:10.1016/j.biocon.2010.09.029.

Lindenmayer, D. B., G. E. Likens, A. Haywood, and L. Miezis. 2011. Adaptive monitoring in the real world: Proof of concept. *Trends in Ecology and Evolution*. doi:10.1016/j.tree.2011.08.002.

Lowe, R. J., A. S. Leon, G. Symonds, J. L. Falter, and R. Gruber. 2015. The intertidal hydraulics of tide-dominated reef platforms. *Journal of Geophysical Research C: Oceans* 120: 4845–4868. doi:10.1002/2015JC010701.

Malaspinas, A. S., Westaway, M., Muller, C., Sousa, V. C., Lao, O., Alves, I. Bergström, A., Athanasiadis, G., et al. 2016. A genomic history of Aboriginal Australia.

Magurran, A. E., S. R. Baillie, S. T. Buckland, J. M. P. Dick, D. A. Elston, E. M. Scott, R. I. Smith, P. J. Somerfield, et al. 2010. Long-term datasets in biodiversity research and monitoring: Assessing change in ecological communities through time. *Trends in Ecology and Evolution* 25: 574–582. doi:10.1016/j.tree.2010.06.016.

Nakashima, D, Krupnik, I, Rubis, JT. (eds) 2018. Indigenous Knowledge for Climate Change Assessment and Adaptation. Cambridge University Press. 298 pp.

Nash, K. L., and N. A. J. Graham. 2016. Ecological indicators for coral reef fisheries management. *Fish and Fisheries* 17: 1029–1054. doi:10.1111/faf.12157.

Nunn, P. D., and N. J. Reid. 2016. Aboriginal Memories of Inundation of the Australian Coast Dating from More than 7000 Years Ago. *Australian Geographer* 47. Taylor & Francis: 11–47. doi:10.1080/00049182.2015.1077539.

Peer, N., E. K. Muhl, J. Janna, M. Brown, S. Zukulu, and P. Mbatha. 2022. Community and Marine Conservation in South Africa: Are We Still Missing the Mark? *Frontiers in Marine Science* 9. doi:10.3389/fmars.2022.884442.

Pople, A. R., S. R. Phinn, N. Menke, G. C. Grigg, H. P. Possingham, and C. McAlpine. 2007. Spatial patterns of kangaroo density across the South Australian pastoral zone over 26 years: Aggregation during drought and suggestions of long distance movement. *Journal of Applied Ecology*. doi:10.1111/j.1365-2664.2007.01344.x.

Purcell, S. W. 2002. Intertidal reefs under extreme tidal flux in Buccaneer Archipelago,Western Australia. *Coral Reefs* 21: 191–192. doi:10.1007/s00338-002-0223-z.

R Core Team. 2021. R: A Language and environment for statistical computing. Edited by R Development Core Team. *R Foundation for Statistical Computing*. R Foundation for Statistical Computing. Vienna, Austria: R Foundation for Statistical Computing.

Rist, P., W. Rassip, D. Yunupingu, J. Wearne, J. Gould, M. Dulfer-Hyams, E. Bock, and D. Smyth. 2019. Indigenous protected areas in Sea Country: Indigenous-driven collaborative marine protected areas in Australia. *Aquatic Conservation: Marine and Freshwater Ecosystems* 29: 138–151. doi:10.1002/aqc.3052.

Ross, H., C. Grant, C. J. Robinson, A. Izurieta, D. Smyth, and P. Rist. 2009. Co-management and indigenous protected areas in Australia: Achievements and ways forward. *Australasian Journal of Environmental Management* 16: 242–252. doi:10.1080/14486563.2009.9725240.

Rouja, P., E. Dewailly, C. Blanchet, and B. Community. 2003. Fat, fishing patterns, and health among the Bardi people of North Western Australia. *Lipids* 38: 399–405.

Sale, P. F. 1978. Coexistence of coral reef fishes - a lottery for living space. *Environmental Biology of Fishes* 3: 85–102. doi:10.1007/BF00006310.

Sale, P. F., and W. A. Douglas. 1984. Temporal Variability in the Community Structure of Fish on Coral Patch Reefs and the Relation of Community Structure to Reef Structure. *Ecology* 65: 409–422. doi:10.2307/1941404.

Schobernd, Z. H., N. M. Bacheler, and P. B. Conn. 2014. Examining the utility of alternative video monitoring metrics for indexing reef fish abundance. *Canadian Journal of Fisheries and Aquatic Sciences* 71: 464–471. doi:10.1139/cjfas-2013-0086.

Sherman, C. S., A. Chin, M. R. Heupel, and C. A. Simpfendorfer. 2018. Are we underestimating elasmobranch abundances on baited remote underwater video systems (BRUVS) using traditional metrics? *Journal of Experimental Marine Biology and Ecology* 503. Elsevier B.V.: 80–85. doi:10.1016/j.jembe.2018.03.002.

Souther S, Colombo S, Lyndon NN, 2023. Integrating traditional ecological knowledge into US public land management: Knowledge gaps and research priorities. *Frontiers in Ecology and Evolution* 11:98812.

Strand, M., N. Rivers, and B. Snow. 2022. Reimagining Ocean Stewardship: Arts-Based Methods to 'Hear' and 'See' Indigenous and Local Knowledge in Ocean Management. *Frontiers in Marine Science* 9: 1–19. doi:10.3389/fmars.2022.886632.

Thibaut, L. M., and S. R. Connolly. 2020. Hierarchical modeling strengthens evidence for density dependence in observational time series of population dynamics. *Ecology* 101: e02893. doi:10.1002/ecy.2893.

Thorburn, D. C., D. L. Morgan, A. J. Rowland, and H. S. Gill. 2007. Freshwater sawfish Pristis microdon Latham, 1794 (Chondrichthyes: Pristidae) in the Kimberley region of Western Australia. *Zootaxa* 1471: 27–41. doi:10.1007/s10641-007-9306-6.

Whitmarsh, S. K., P. G. Fairweather, and C. Huveneers. 2017. What is Big BRUVver up to? Methods and uses of baited underwater video. *Reviews in Fish Biology and Fisheries* 27. Springer International Publishing: 53–73. doi:10.1007/s11160-016-9450-1.

Willis, T., and R. Babcock. 2000. A baited underwater video system for the determination of relative density of carnivorous reef fish. *Marine and Freshwater Research* 51: 755–763.

Wilson, S. K., N. A. J. Graham, M. S. Pratchett, G. P. Jones, and N. V. C. Polunin. 2006. Multiple disturbances and the global degradation of coral reefs: are reef fishes at risk or resilient? *Global Change Biology* 12: 2220–2234.

Wilson, S. K., R. C. Babcock, R. Fisher, T. H. Holmes, J. A. Y. Moore, and D. P. Thomson. 2012. Relative and combined effects of habitat and fishing on reef fish communities across a limited fishing gradient at Ningaloo. *Marine Environmental Research* 81: 1–11.

Supplementary Information

Title: **Communicating uncertainty in Indigenous sea Country monitoring with Bayesian statistics: towards more informed decision-making**

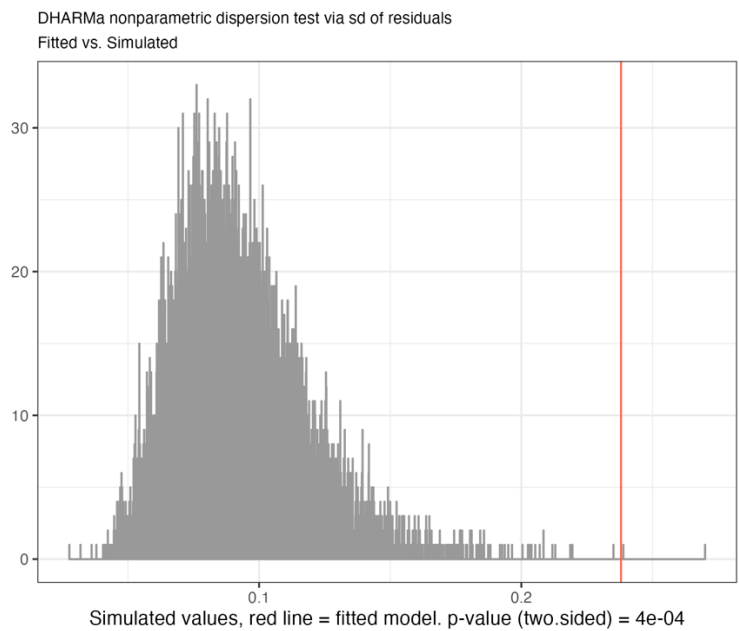

**Fig. S1.** Non-parametric test for dispersion from posterior simulated residuals of model in equation (1) refitted assuming a Poisson distribution instead of a Negative Binomial. Model validation plot generated using the R package DHARMa version 0.4.5 (Hartig 2022).

Hartig, F (2022). DHARMa: Residual Diagnostics for Hierarchical (Multi-Level / Mixed) Regression Models. R package version 0.4.5. https://CRAN.R-project.org/package=DHARMa

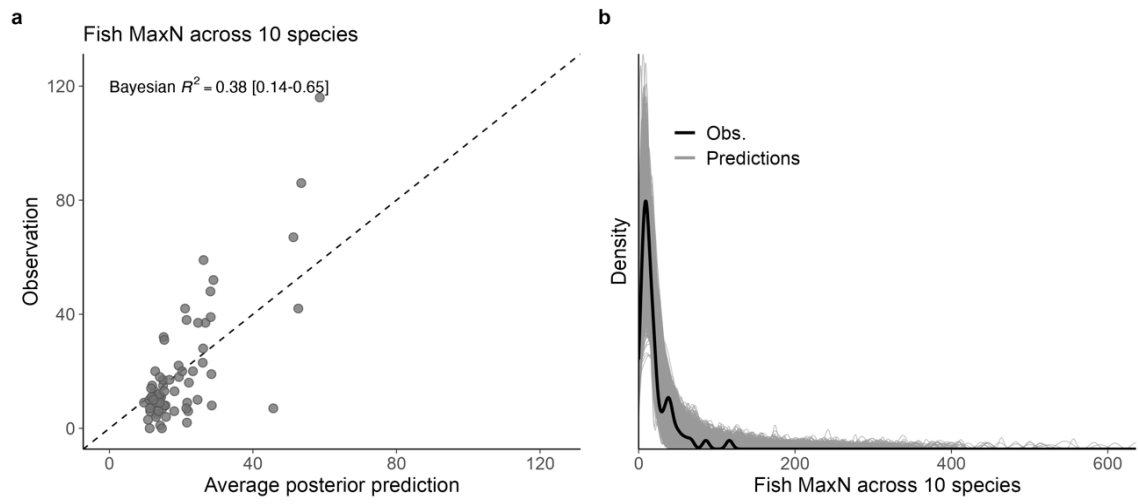

**Fig. S2.** Posterior predictive checks for main model of equation (1). (a) depicts the relationship between observations and average predictions of the *MaxN* summed across the ten fish species in Table 1. Dashed line represents a 1:1 relationship. In (b), observed density of *MaxN* (thick black line) overlaying 1,000 mean posterior predictions (thin grey lines).